\documentclass{aa}  

\usepackage{graphicx}
\usepackage{txfonts}
\usepackage{graphicx}	
\usepackage{amsmath}	
\usepackage{amssymb}	
\usepackage{hyperref}
\usepackage{color}
\usepackage{ulem}
\usepackage{CJK}
\begin{document}
\begin{CJK*}{UTF8}{gbsn}

    \title{New method for estimating molecular cloud distances based on \textit{Gaia}, 2MASS, and the TRILEGAL galaxy model}
    
    \author{Juan Mei (梅娟)
          \inst{1,2,3}
          \and
          Zhiwei Chen (陈志维) \inst{3}\thanks{Corresponding author: Zhiwei Chen}
          \and
          Zhibo Jiang (江治波) \inst{2,3,4}
          \and
          Sheng Zheng (郑胜) \inst{1,2} 
          \and   \\
          Haoran Feng (冯浩然) \inst{3,4}
          }
    
    \institute{Center for Astronomy and Space Sciences, China Three Gorges University, 8 University Road, 443002 Yichang, China
        \and
             College of Science, China Three Gorges University, 8 University Road, 443002 Yichang, China
        \and
             Purple Mountain Observatory, Chinese Academy of Sciences, 10 Yuanhua Road, 210023 Nanjing, China\\
             \email{zwchen@pmo.ac.cn}
        \and
            University of Science and Technology of China, Chinese Academy of Sciences, Hefei 230026, China
             }
    
    \date{}
\titlerunning{Distances of molecular clouds}
\authorrunning{Mei et al.}

\abstract
{
We propose a new method for estimating the distances of molecular clouds traced by CO line emission. Stars from 2MASS and \textit{Gaia} EDR3 are selected as on-cloud stars when they are projected on a cloud. The background on-cloud stars have redder colors on average than the foreground stars. Instead of searching for stars projected away from the cloud, we employed the TRILEGA galaxy model to mimic the stellar population without cloud extinction along the sightline toward the cloud. Our method does not require an exact boundary of a cloud. The boundaries are highly variable and depend on the sensitivity of the molecular line data. For each cloud, we compared the distributions of on-cloud stars to the TRILEGAL stellar populations in the diagram of $J-K_s$ color versus distance. The intrinsic $J-K_s$ colors of main-sequence and evolved stars from TRILEGAL were considered separately, and they were used as the baseline for subtracting the observed $J-K_s$ colors. The baseline-corrected $J-K_s$ color was deployed with the Bayesian analysis and Markov chain Monte Carlo sampling to determine the distance at which the $J-K_s$ color jump is largest. This method was successfully applied to measure the distances of 27 molecular clouds, which were selected from previously published cloud samples. By replacing TRILEGAL with the GALAXIA galaxy model, we were able to measure the distances for 21 of the 27 clouds. The distances of the 21 clouds based on the GALAXIA model agree well with those based on the TRILEGAL model. The distances of the 27 clouds estimated by this method are consistent with previous estimates. We will apply this new method to a larger region of the gaseous galactic plane, in particular, for the inner galactic region, where a region free of CO emission is hard to separate from the crowded field of clouds.

}

\maketitle
%

\section{Introduction}
Molecular clouds are the birthplaces of stars. The distance is an important parameter of a molecular cloud when the intrinsic physical properties (mass and size) are to be estimated. Determining the distances of molecular clouds is crucial for characterizing the initial conditions of star formation \citep{McKee+2007, Kennicutt+2012, HeyerDame+2015} and the gaseous spiral arms of the Milky Way \citep{Dame+2001, Xu+2018}.

It is always a challenging task to determine the distances of molecular clouds. Several methods were applied to estimate the distances of molecular clouds. The distances of molecular clouds can be estimated from their radial velocities ($V_\mathrm{lsr}$), which are assumed to be attributed to the large-scale rotation of the spiral arms of the Milky Way. The distances estimated from $V_\mathrm{lsr}$ of molecular clouds are commonly referred to as kinematic distances \citep{Roman-Duval+2009, Reid+2014}. The kinematic distances have large uncertainties and ambiguity problems, especially in the inner Galaxy. The cloud distances can also be estimated by identifying objects associated with or within the molecular cloud, such as OB-associations, young open clusters, \ion{H}{ii} regions, young stellar objects (YSOs), and masers, whose distances are measured from the trigonometric parallaxes or photometry of stars \citep{Xu+2006, Russeil+2007, Reid+2019, Cantat-Gaudin+2021, Marton+2022, Zhang+2023}. The trigonometric parallaxes measured by \textit{Gaia} \citep{gaia+2016} for stars up to several kiloparsec provide another approach for estimating the distances of molecular clouds by locating the extinction break point along the line of sight (LOS) toward molecular clouds.

The extinction method relies on accurately estimating the distance and extinction of numerous stars. \citet{Green+2014} presented a method for deriving reddening and distances of stars from Panoramic Survey Telescope and Rapid Response System (PanSTARRS-1) photometry and produced a three-dimensional (3D) dust map. Based on the technique of \citet{Green+2014}, \citet{Schlafly+2014} simultaneously inferred the reddening and distances of stars. They measured the distances of molecular clouds selected from the catalog of \citet{Magnani+1985} by their extinction break point. The release of astrometric data from \textit{Gaia} has brought about significant changes in measuring the distance, as it provides a more precise stellar parallax. Combined with \textit{Gaia} data, several studies presented new 3D extinction maps and have estimated the extinction and distances for millions of stars \citep{Green+2019, ChenBQ+2019, Lallement+2019, Guo+2021, Sun+2021a}. One approach involves modeling the dust extinction profiles along different LOS, where the measured cloud is searched for dust clouds from 3D dust structures \citep{ChenBQ+2020a, Guo+2022}. Another approach involves searching for the location of the extinction break point that is caused by molecular clouds \citep{Yan+2019, Zucker+2019, Zucker+2020, ChenBQ+2020b, Sun+2021b}. It can also be applied to supernova remnants \citep{Shan+2019, Wang+2020, Zhao+2020, Leike+2021}.

In this work, we make a new attempt to estimate the  molecular cloud distances. Principally, stars in front of a molecular cloud have bluer colors on average than those behind a cloud. Instead of searching for stars around the molecular cloud (off-cloud stars), we use the TRILEGAL galaxy model \citep{Girardi+2016} to mimic the stellar population without cloud extinction along the LOS toward the cloud. This means that our method does not depend strongly on the exact boundary of a cloud. The TRILEGAL model is able to simulate stellar populations along sightlines. We fit the baseline separately for main-sequence (MS) and red giant stars (RGs) to calibrate the corresponding stellar colors. We have estimated the distances of measured molecular clouds from \citet{Yan+2021} to confirm the reliability of the method.

The paper is structured as follows. In Section 2 we describe the data. In Section 3 we introduce our method for estimating the distances of molecular clouds. We present our results and discussions in Section 4. We summarize in Section 5.

\section{Data}
Although the bulk masses of molecular clouds are contributed by molecular hydrogen, molecular clouds are well traced by the $^{12}$CO molecule, which is the most abundant molecule in molecular clouds after molecular hydrogen. The $J=1-0$ transition line emission of $^{12}$CO can be observed from ground-based radio telescopes through the atmospheric window at the 3 millimeter wavelength \citep{Dame+2001}. In contrast to the $^{13}$CO and C$^{18}$O $J=1-0$ emission, $^{12}$CO $J=1-0$ emission is extended and strong, and it is therefore often chosen to construct the inventory of molecular clouds. We used the $^{12}$CO $J=1-0$ line emission data from the Milky Way Imaging Scroll Painting (MWISP \footnote{\url{http://www.radioast.nsdc.cn/mwisp.php}}) survey by the Purple Mountain Observatory (PMO) 13.7 m millimeter telescope, located at Delingha in China \citep[see more details of MWISP in][]{Su+2019}. The rms noise of the $^{12}$CO emission of the MWISP survey is about 0.5 K at a velocity resolution of $0.16\,\mathrm{km^{-1}}$ and a grid spacing of $30\arcsec\times30\arcsec$. The half-power beamwidth (HPBW) of the 13.7 m telescope at 115 GHz is $50\arcsec$.

Molecular clouds are identified from the $^{12}$CO $J=1-0$ emission using the DBSCAN algorithm \citep{Ester+1996}. \citet{Yan+2021} applied DBSCAN to identify molecular clouds from the MWISP $^{12}$CO $J=1-0$ line emission between $l=25\degr-50\degr$ and in the velocity range -6 to $30\,\mathrm{km\,s^{-1}}$. By setting the two DBSCAN parameters Eps=1 and MinPts=4, they extracted 359 molecular clouds and measured distances for 27 molecular clouds using the background-eliminated extinction-parallax II (BEEP-II) method. For simplicity, we chose 27 molecular clouds with distance estimates from the cloud sample in \citet{Yan+2021} and used the cloud G045.1-03.4 as an example to explain our method, which we present in this work. 

We used the astrometric catalog \textit{Gaia} Early Release 3 (\textit{Gaia} EDR3; \citealt{Gaia3+2021}) and the near-IR photometry catalog Two Micron All Sky Survey (2MASS; \citealt{Skrutskie+2MASS+2006}). The \textit{Gaia} EDR3 consists of astrometry and photometry for 1.8 billion sources, which improves the parallax precision by 30$\%$ over that of \textit{Gaia} Data Release 2 (\textit{Gaia} DR2; \citealt{gaia2+2018}). The adopted stellar distances were estimated by \citet{BailerJ+2021}, who transferred the \textit{Gaia} EDR3 parallaxes to distances using a Bayesian approach. The 2MASS was conducted with two 1.3 m telescopes to provide full sky coverage. It has three near-IR bands, $J$, $H$, and $K_s$, centered at 1.25, 1.65, and 2.16 $\mu m$, respectively. The systematic uncertainties of 2MASS photometric measurements are $<$0.03 mag. 

We merged the \textit{Gaia} and 2MASS catalogs using a cross-match radius of $1\arcsec$ and took the $J-K{_s}$ color from the 2MASS catalog \citep{Cutri+2MASScatalog+2003} and the distance from \citet{BailerJ+2021}. The uncertainty of the stellar distance, $\Delta d$, is defined as
\begin{eqnarray}
	\Delta d=\frac{1}{2}(d_{\rm upper}-d_{\rm lower}),
\end{eqnarray}
where $d_{\rm upper}$ and $d_{\rm lower}$ are the 16th and 84th percentiles of the distance given by \citet{BailerJ+2021}. The uncertainty of the $J-K_s$ color, $\Delta (J-K{_s})$, was propagated from the photometric uncertainties in the $J$ and $K_s$ bands. 

\section{Method} 
The light passing through a molecular cloud is attenuated by dust within the cloud. The average color of the background stars behind a molecular cloud is redder than that of the foreground stars. The basic idea is to compare the colors of background stars to those of foreground stars, and to locate the distance at which the color difference due to cloud extinction is most prominent. This distance is regarded as the distance to the molecular cloud. This approach greatly benefits from the precise trigonometric parallaxes of billions of stars measured by \textit{Gaia}.  

\subsection{Selection of on-cloud stars}
\begin{figure*}
    \centering
	\includegraphics[width=0.85\linewidth]{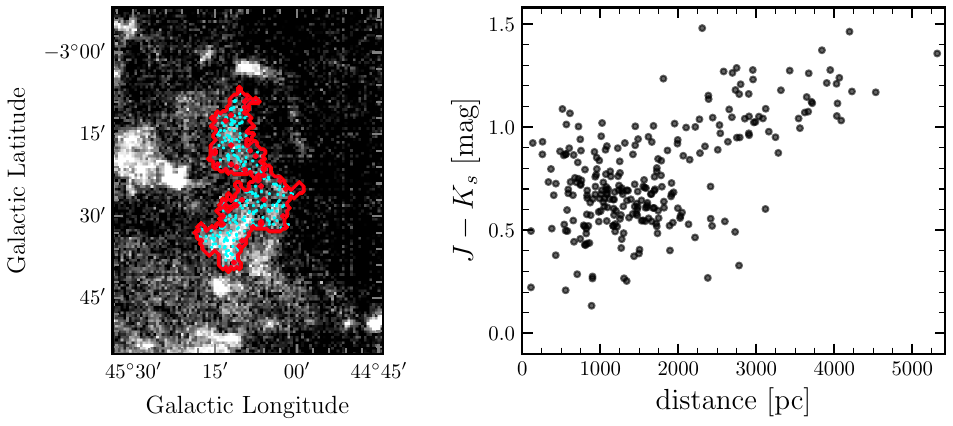}
    \caption{Left panel: Selected on-cloud stars for G045.1-03.4. The red contour shows the molecular cloud footprint, and the cyan points represent on-cloud stars. Right panel: $J-K{_s}$ color vs. distance diagram. It has two branches.}
    \label{onstars}
\end{figure*}

For each of the 27 clouds in this work, we searched for stars from the merged \textit{Gaia}-2MASS catalog based on the position and size of the cloud. A large number of stars can be found within a certain radius for each cloud. We selected the stars projected on the cloud, called on-cloud stars. There are three criteria for selecting the on-cloud stars: (1) The stars lie within the irregular boundary of the molecular cloud. (2) The relative uncertainty of the stellar distance is lower than 10$\%$, that is, $\Delta d/d$ $\leq$ 0.1. (3) For each cloud, we calculated a mean rms noise, $\sigma_{\rm int}$, of integrated CO emission ($W_{\rm CO}$), based on the $^{12}$CO spectra of a cloud in a certain velocity range. We set a threshold $\mathrm{CO}_{\rm cut}=5\sigma_{\rm int}$. On-cloud stars falling within areas with $W_{\rm CO} \geq \mathrm{CO}_{\rm cut}$ were kept. The above criteria can be adjusted to optimize the sample size of selected on-cloud stars for each of the 27 clouds. We can increase the CO$_{\rm cut}$ to enhance the contrast of $J-K_s$ colors between foreground and background stars. When the selected on-cloud stars for a cloud were not sufficient, we decreased the CO$_{\rm cut}$ to gather more on-cloud stars. As an example shown in the left panel of Fig.~\ref{onstars}, the selected on-cloud stars (cyan points) for the G045.1-03.4 cloud are overlaid on the integrated-intensity map of the $^{12}$CO emission for the cloud. The right panel of Fig.~\ref{onstars} presents the diagram of the $J-K_s$ color versus stellar distance of the selected on-cloud stars for the G045.1-03.4 cloud. In the color-distance diagram, stellar colors start to split into two trends at distance $\gtrsim 1$ kpc. We examine the two trends below, in combination with synthetic stellar population. 

\subsection{Synthesized stellar population by the TRILEGAL galaxy model}
The distributions of on-cloud stars in the color-distance diagram were compared with the synthetic stellar population, which was adopted from the TRILEGAL galaxy model \citep{Girardi+2016}. \citet{Chenzw+2022} compared the $J-K_s$ colors of stars toward RCW\,120 to the synthetic stellar population of the TRILEGAL galaxy model. Stars with distances $\lesssim 1.5$ kpc match the TRILEGAL stellar population in the color-distance diagram well \citep[c.f. fig.3 in][]{Chenzw+2022}. For each cloud, we generated a collection of synthetic stellar populations within an area slightly larger than the cloud from the online input form of TRILEGAL 1.6 \footnote{\url{http://stev.oapd.inaf.it/cgi-bin/trilegal}}. We tweaked the options provided in the online input form of TRILEGAL 1.6. We chose the binary-corrected initial mass function (IMF) of single stars with a break point at $0.5\,M_\odot$ \citep{Kroupa+2002}. The photometry system was set to 2MASS $JHK_s$. For each cloud, we adjusted the limiting magnitude in the $K_s$ band by directly counting the $K_s$-band magnitudes of the on-cloud stars. The distance modulus resolution of the Galaxy components is 0.05 mag. The online input form of TRILEGAL 1.6 provides the option for the dust extinction attributed to the diffuse interstellar medium (ISM) along the sight line. We selected several areas free of $^{12}$CO $J=1-0$ line emission to simulate the diffuse ISM, and adjusted the $K_s$-band limiting magnitudes for these areas. We find a good match in the color-distance diagram between the 2MASS+Gaia combination toward these cloud-free areas and the TRILEGAL stellar population with the dust extinction trend of $A_V/d = 1\,\mathrm{mag\,kpc}^{-1}$. The 1$\sigma$ dispersion of the extinction is 0.01 times the total extinction. The Sun is located at a galactocentric distance of $8700\, \mathrm{pc}$ and at a height of $24.2\, \mathrm{pc}$ along the galactic disk. For the thin disk, we chose a squared hyperbolic secant along $z$, $sech^2(0.5z/h_{z,d})$, where $h_{z,d}$ increases with age $t$ ($h_{z,d}=z_0(1+t/t_0)^\alpha$). The thick disk, similar to the thin disk, is described by a sech$^2$ function, $sech^2(0.5z/h_{z,td})$. The halo is represented by an oblate spheroid with a $r^{1/4}$ density profile, and the bulge is modeled as a triaxial structure. The output parameters of TRILEGAL mainly include the Galactic component (selecting the thin disk is sufficient), age,  luminosity, effective temperature, surface gravity ($\log g$), absolute distance modulus, apparent magnitude in desired filters, and mass.

\begin{figure*}
    \centering
	\includegraphics[width=0.92\linewidth]{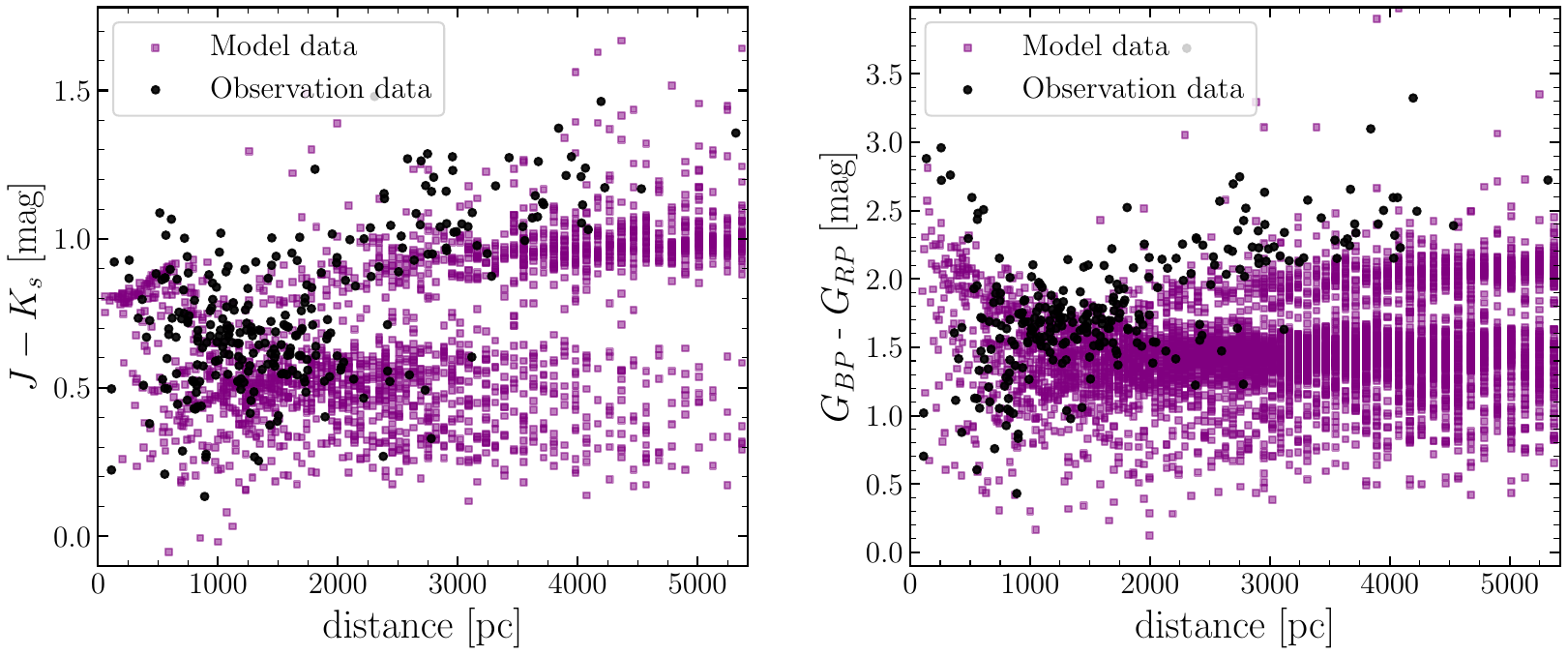}
	\caption{Left and right panels: $J-K_s$ color vs. distance and $G_{BP}-G_{RP}$ color vs. distance diagram for G045.1-03.4, respectively. The purple squares represent the model data, and the black points represent the observation data.}
	\label{color-distance}
\end{figure*}

The TRILEGAL stellar population generated with the above tweaks populate the color-distance diagram similarly to the 2MASS+Gaia combination, as shown in the left panel of Fig.~\ref{color-distance}, for instance. As a comparison, the right panel of Fig.~\ref{color-distance} presents the same stellar populations in the \textit{Gaia} $G_{BP}-G_{RP}$ versus distance diagram. The TRILEGAL stellar population generally reproduces the observations in both $J-K_s$ and $G_{BP}-G_{RP}$ colors. However, TRILEGAL predicts stellar colors that are more compatible with the 2MASS photometric data than with the \textit{Gaia} photometric data. The $G_{BP}-G_{RP}$ colors of TRILEGAL are lower than the observed values by \textit{Gaia} for late-type dwarf stars, which are at closer distances with intrinsically red colors. The TRILEGAL model adopts $R_V=3.1$ as the default extinction law. The $G_{BP}-G_{RP}$ colors are more sensitive to dust extinction than the $J-K_s$ colors. Any differential from the nominal $R_V=3.1$ extinction law might lead to an observable mismatch between the observed and synthetic $G_{BP}-G_{RP}$ colors. A detailed comparison between TRILEGAL model and various photometric surveys is beyond the scope of this work. For simplicity, we use the 2MASS+Gaia combination in the following analysis. 

\begin{figure}
    \centering
	\includegraphics[width=1\linewidth]{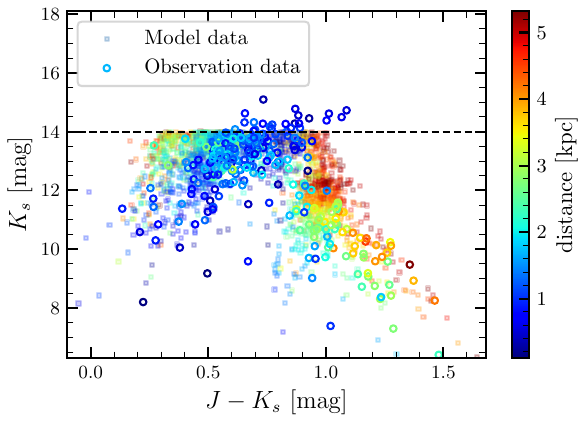} 
	\caption{ $K_s$ vs. $J-K_s$ diagram of the observation and model data for G045.1-03.4. The squares represent the model data, the hollow circles represent the observation data, and the color represents the distance. The dashed black line is $K_s$ = 14 mag.}
	\label{magnitude-distance-Ks}
\end{figure}

Figure~\ref{magnitude-distance-Ks} compares the TRILEGAL stellar population with the 2MASS+Gaia combination in the $K_s$ versus $J-K_s$ diagram for G045.1-03.4. The $K_s$-band limiting magnitude was set to 14 mag for TRILEGAL to generate the synthesized stellar population. Fewer than 10\% of the observed stars have a $K_s$ magnitude fainter than 14 mag, corresponding to the few stars lying above the truncation of the $K_s$ magnitude for the synthesized stellar population. These faint stars are at distances closer than about 1.5 kpc and are intrinsically red with $J-K_s>0.6$, suggesting that they are cool dwarf stars. For most of the observed stars, the synthesized stellar population matches the observations in the $K_s$ versus $J-K_s$ diagram well. The synthesized and observed populations both present two dominant branches that split at $J-K_s\approx0.7-0.8$. We argue that the TRILEGAL galaxy model can be used to generate a synthesized stellar population along the LOS toward molecular clouds, but without cloud extinction. We computed a TRILEGAL stellar population for each of the 27 molecular clouds. Instead of searching off-cloud stars, we used the $J-K_s$ color and distance of each synthesized star to simulate the baseline attributed to the diffuse ISM in the color-distance diagram.

\begin{figure} 
    \centering
	\includegraphics[width=1\linewidth]{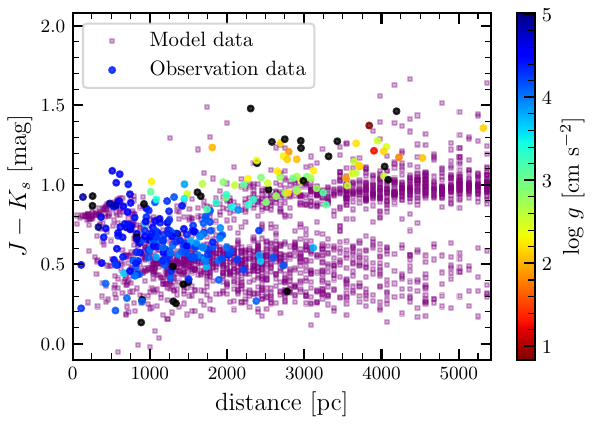}
	\caption{Distributions of the observation and model data in the color-distance diagram for G045.1-03.4. The purple squares represent the model data, the colored points represent the observation data with log $g$, and the black points represent the observation data without log $g$.}
	\label{color-distance_logg}
\end{figure}

\subsection{Elimination of background}
The observed and synthesized stellar populations both start to split into two branches at a distance of $\gtrsim 1\,\mathrm{kpc}$ in the color-distance diagram, as shown in Fig.~\ref{color-distance_logg}. The synthesized stellar population at a distance of $\gtrsim1\,\mathrm{kpc}$ with $J-K_s>0.8$ is dominated by RGs with $\log g<3.45$, while MSs are generally of bluer colors $J-K_s<0.8$ at the same distances. By adopting the $\log g$ estimates from \textit{Gaia} DR3 \citep{GaiaDR3+2023,Andrae+2023}, the observed stars with \textit{Gaia} DR3 $\log g$ estimates are displayed with different colors in Fig.~\ref{color-distance_logg}. The $\log g$ distributions of the observed stars match the TRILEGAL predictions well. MSs of higher $\log g$ are distinct from RGs of lower $\log g$ in the color-distance diagram. Cool MSs with  $0.7<J-K_s<1.1$ are usually faint sources that are only visible at distances $<1\,\mathrm{kpc}$. As the distance increases, cool MSs become very faint and escape detection by 2MASS, but MSs of earlier types remain detectable. This distance dilution effect is seen for the observed and synthesized populations, that is, the $J-K_s$ colors of MSs become bluer with increasing distance \citep[see also][]{Chenzw+2022}.

\begin{figure*}
    \centering
    \includegraphics[width=0.92\linewidth]{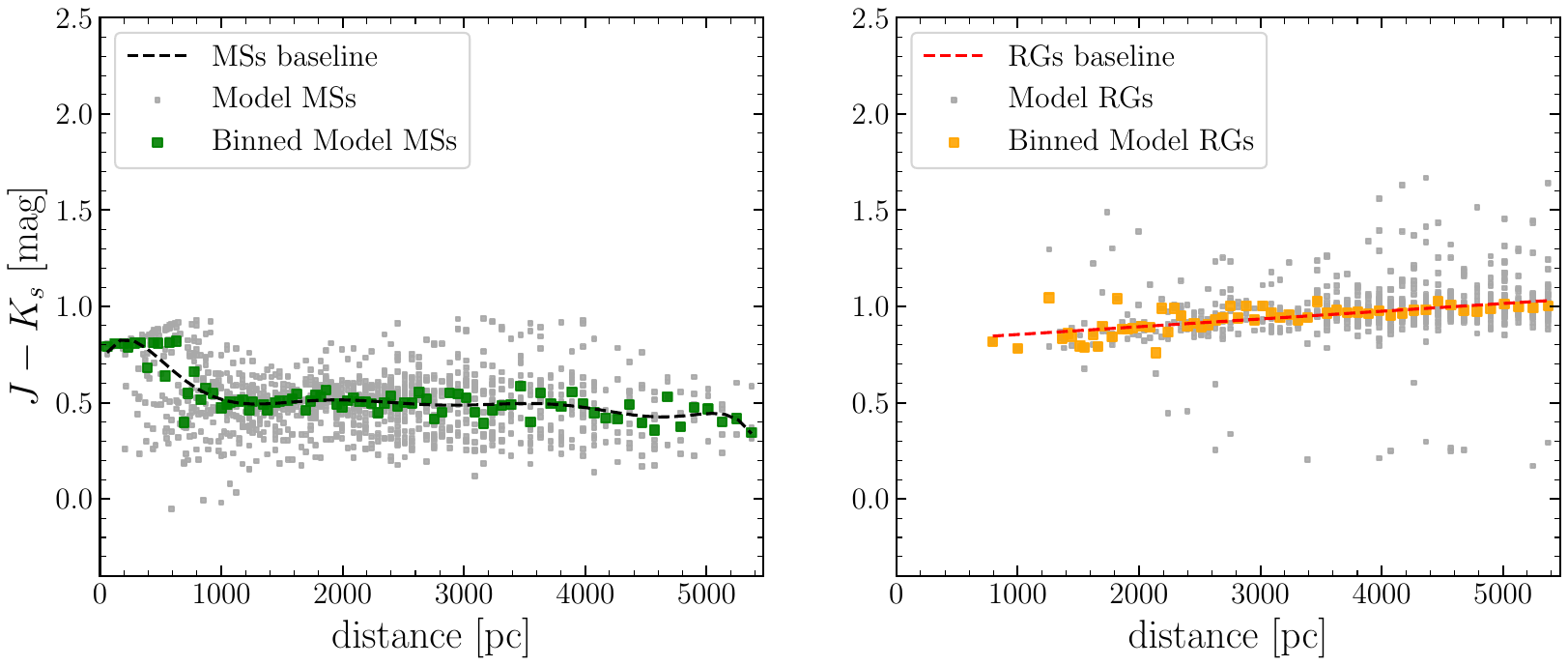}
    \caption{Left and right panels: Fit baselines for MSs (dashed black line) and for RGs (dashed red line) in the model data, respectively. The gray squares present the model data. The green and orange squares present the binned MSs and RGs stars (every 50 pc), respectively.}
	\label{baselines}
\end{figure*}

We classified the TRILEGAL stellar population into MSs and RGs based on the modeled $\log g$ of each simulated star. For simplicity, MSs were defined to have $\log g>3.45$, and RGs have $\log g<3.45$. The MSs and RGs samples of TRILEGAL were separately as depicted in the left and right panels of Fig.~\ref{baselines}, respectively. We chose a bin of 50 pc to obtain the median colors of MSs and RGs in each distance bin. This is shown as the green and orange squares in the left and right panels of Fig.~\ref{baselines}, respectively. We further applied a polynomial fit to the median colors of MSs and a linear fit to the median colors of RGs at various distances, and we obtained the two color baselines shown in Fig.~\ref{baselines}: the MSs and the RGs baselines. 

Different from the TRILEGAL model without cloud extinction, the observed stars that are located behind molecular clouds are reddened due to cloud extinction along the LOS. Cloud extinction leads to higher $J-K_s$ colors for MSs, resulting in the degeneracy between reddened MSs and RGs on colors. Although \textit{Gaia} DR3 estimated $\log g$ for numerous stars from their low-resolution spectra in the \textit{Gaia} $G_{BP}$ and $G_{RP}$ bands, the low-resolution spectra of reddened MSs are likely analogous to those of the RGs. The degeneracy between reddened MSs and RGs still remains. The stars at a $distance>1\,\mathrm{kpc}$ of with $J-K_s>0.8$ might originally be RGs or MSs that are reddened by a certain amount of cloud extinction. For example, late-F MSs with intrinsic colors of $J-K_s\approx0.3$ are still detectable by 2MASS and \textit{Gaia}, even when they are reddened to $J-K_s\sim0.8-1.0$ by cloud extinction of $A_V\sim2.5-3.5$. The 27 clouds studied in this work are generally not dense; otherwise, we expect very few background stars with a \textit{Gaia} detection. 

\begin{figure*}
    \centering
	\includegraphics[width=0.92\linewidth]{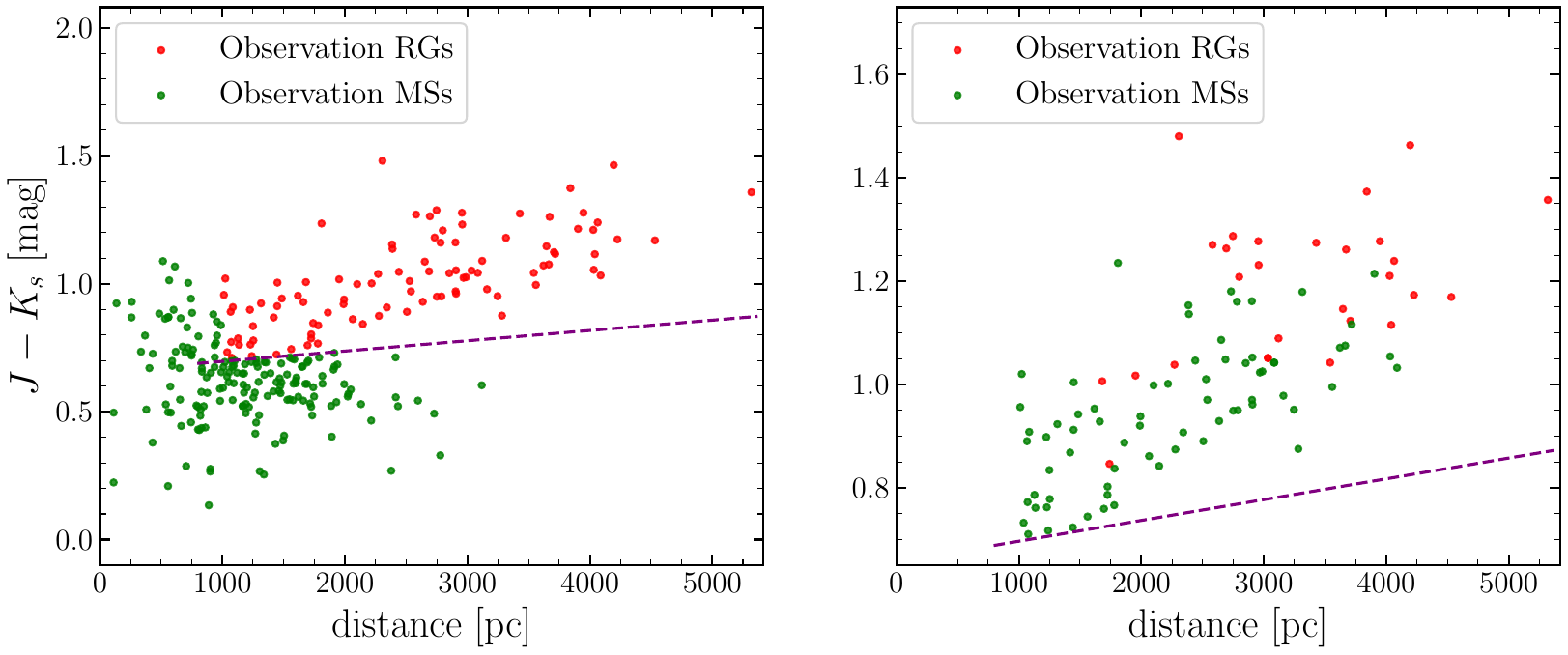}
	\caption{Left panel: Distinguished MSs and RGs in the observation data using the 3$\sigma$ lower bound line of the model-fit RGs baseline (dashed purple line) and a distance of 1 kpc. Right panel: Further distinction between the RGs and reddened stars for the differentiated RGs, and reddened stars are classified as MSs. The green and red points show the MSs and RGs of the observation data, respectively.}
	\label{distinguished-observation}
\end{figure*}

\begin{figure}
    \centering
	\includegraphics[width=0.92\linewidth]{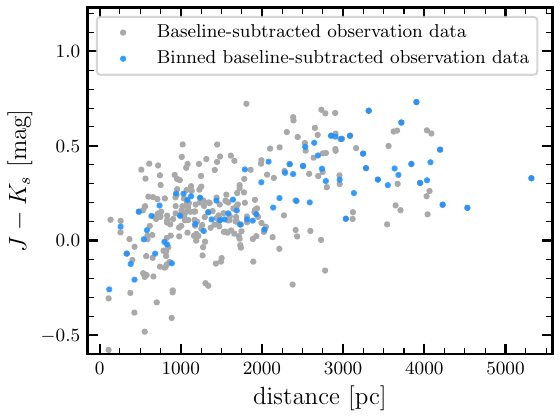}
	\caption{Distribution of the observation data after the color is calibrated. The gray points present the baseline-subtracted observation data. The blue points plot the mean color of the binned stars (every 50 pc for the distance), which are only used to confirm the results for inspection by eye and not to calculate the distance.}
	\label{baseline-subtracted}
\end{figure}

The stars with distances greater than 1 kpc and colors exceeding the 3$\sigma$ lower bound line of RGs baseline are more likely to be RGs, which are marked as red dots in the left panel of Fig.~\ref{distinguished-observation}. Considering the degeneracy between reddened MSs and RGs on colors, we binned every 100 pc for the observed and synthesized populations simultaneously, and we sorted the observed colors for each distance bin. We further distinguished reddened MSs and RGs according to the fractions of MSs and RGs in TRILEGAL, as illustrated in the right panel of Fig.~\ref{distinguished-observation}. We subtracted the corresponding model-fit baseline from (reddened) MSs and RGs of the observation data. Figure \ref{baseline-subtracted} displays the distribution of baseline-subtracted data. The blue points (binned every 50 pc) are only used for inspection by eyes and not to estimate the distance.
 
\subsection{Bayesian analysis and Markov chain Monte Carlo sampling}
We adopted a Bayesian analysis and Markov chain Monte Carlo (MCMC) sampling to derive the molecular cloud distances from the baseline-subtracted data. After subtracting the baseline, the color distribution of on-cloud stars became regular and approximately Gaussian. The stellar reddening caused by the molecular cloud remained. In order to estimate the molecular cloud distance, the Bayesian model used in \citet{Yan+2019} was adopted. We assumed that the distribution of diffuse dust in the on-cloud region is approximately uniform and that a star is either in front of or behind a molecular cloud. Given a molecular cloud distance ($D$), the probability that the star is a foreground star is
\begin{eqnarray}
	f_i=\phi \left(\frac{D-d_i}{\Delta d_i} \right),
\end{eqnarray}
where $\phi(x)=\frac{1}{\sqrt{2\pi}} \int_{\infty}^{x} e^{-t^2/2}\,dt$ is the cumulative distribution function (CDF) of a Gaussian function, and $d_i$ and $\Delta d_i$ are the on-cloud star distance and its standard deviation, respectively. Then, the probability that the star is a background star is (1 - $f_i$). The probability density function (PDF) of the Gaussian distribution with a mean $\mu$ and a standard deviation $\sigma$ is
\begin{eqnarray}
	p(J-K_s|\mu,\sigma)= \frac{1}{\sqrt{2\pi} \sigma} \exp{\left(-\frac{1}{2} \left(\frac{(J-K_s)-\mu}{\sigma} \right)^2  \right)}. 
\end{eqnarray}
The likelihoods of the foreground and background stars are denoted as
\begin{eqnarray}
    PF_i =p \left((J-K_s)_i|\mu_1,\sqrt{\sigma_1^2+\Delta(J-K_s)_i^2}  \right),
\end{eqnarray}
\begin{eqnarray}
    PB_i =p \left((J-K_s)_i|\mu_2,\sqrt{\sigma_2^2+\Delta(J-K_s)_i^2} \right).
\end{eqnarray}
Therefore, the likelihood of a star is
\begin{eqnarray}
	p \left((J-K_s)_i|D,\mu_1,\sigma_1,\mu_2,\sigma_2 \right)=f_i PF_i+(1-f_i)PB_i.
\end{eqnarray}
The total likelihood is the product of $N$ on-cloud star likelihoods, that is, 
\begin{eqnarray}
	L=\prod\limits_{i=0}\limits^{N} p_i.
\end{eqnarray}

The Bayesian model contains five parameters: the molecular cloud distance ($D$), the mean ($\mu_1$) and standard deviation ($\sigma_1$) of the foreground star colors, and the mean ($\mu_2$) and standard deviation ($\sigma_2$) of the background star colors. For each molecular cloud, we assumed that the prior distribution of five parameters ($D$, $\mu_1$, $\sigma_1$, $\mu_2$, and $\sigma_2$) was a uniform distribution, denoting the minimum and maximum of the on-cloud stars as $D_{\rm min}$ and $D_{\rm max}$. To prevent the MCMC process from considering all stars as background stars for convergence to small molecular cloud distances, we adjusted the distance minimum to be located 200 pc after the nearest star distance, that is, $D \sim U[D_{\rm min}+200, D_{\rm max}]$, where $U$ represents a uniform distribution. Moreover, we required that the mean value $\mu_1$ of the foreground star colors must be lower than the mean value $\mu_2$ of the background star colors and that the standard deviations ($\sigma_1$ and $\sigma_2$) of the foreground and background stars were greater than 0, that is, 

\begin{eqnarray}
    P \left(\mu_1,\sigma_1,\mu_2,\sigma_2 \right)  =
    \begin{cases}
        1 \qquad if  
        \begin{cases} 
            \mu_1 < \mu_2, \\
            \sigma_1 > 0, \\ 
            \sigma_2 > 0,  \\
        \end{cases} \\
        0 \qquad else. 
    \end{cases}
\end{eqnarray}

\begin{figure*}
    \centering
	\includegraphics[width=1\linewidth]{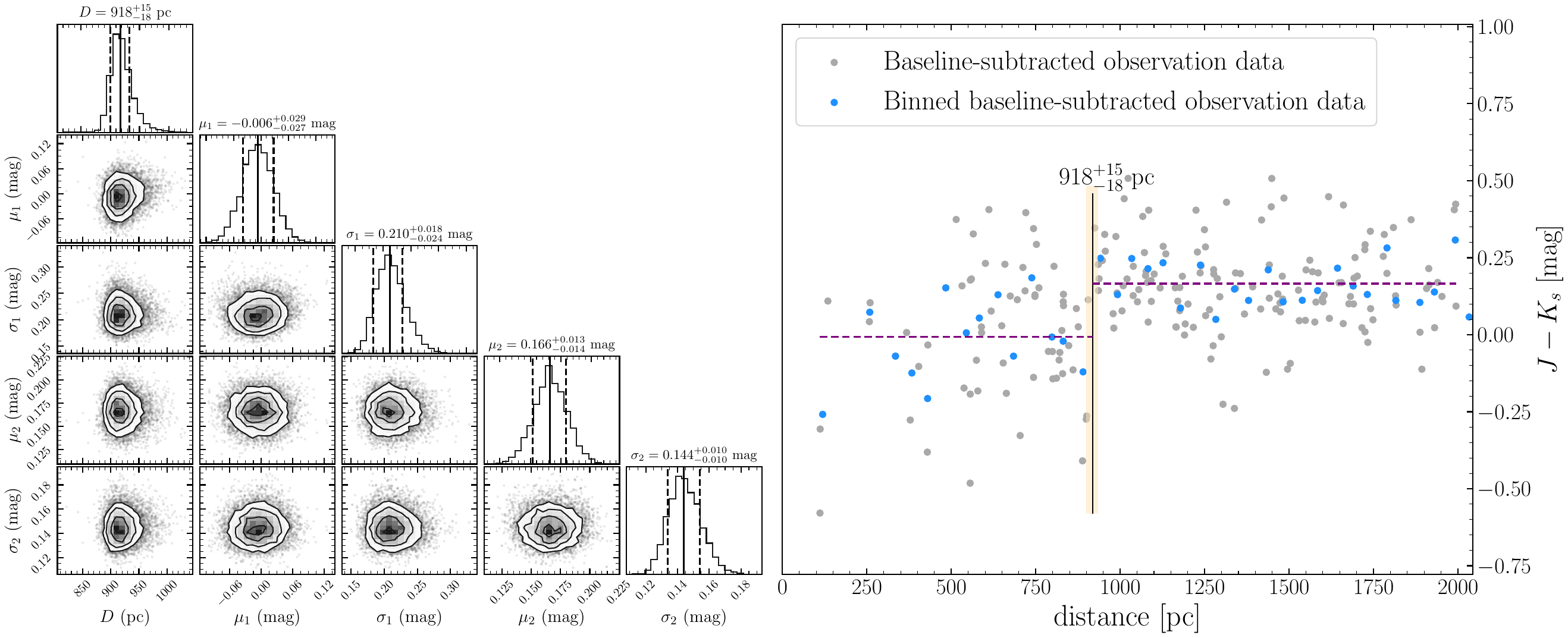}
	\caption{Distances of G045.1-03.4. The corner plots of the MCMC samples of five modeled parameters and their uncertainties are displayed on the left. The right panel displays the distance result. The vertical black lines indicate the estimated distance ($D$), and the shadow areas depict the 16th and 84th percentile values of the cloud distance. The broken horizontal dashed purple line is the color variation of the on-cloud stars after the baseline was subtracted.}
	\label{distance-mcmc}
\end{figure*}

We used the MCMC algorithm \citep{ForemanM+2013} EMCEE package to derive the parameters and uncertainties. For each parameter, we adopted 10000 steps with 10 walkers to run. The last 1000 steps from each walker were used to sample the final posterior. The results for each parameter approximate a Gaussian distribution. We took the 50th percentile value of the posterior distribution as the best estimate of the parameter, with the uncertainties derived from the 16th and 84th percentile values. To avoid the influence of farther molecular clouds, we set a maximum distance ($D_{\rm cut}$) for each molecular cloud. This value was set before estimating clouds distances, which is equivalent to a recursive process. We determined it by eye from the baseline-subtracted data and adjusted it by the near or far distance of the measured molecular cloud. As demonstrated in Fig.~\ref{distance-mcmc}, we display the estimated distance for G045.1-03.4 based on the MCMC sampling. The corner plots show the five parameters $D$, $\mu_1$, $\sigma_1$, $\mu_2$, and $\sigma_2$, together with their 50th, 16th, and 84th percentile. The distance of the G045.1-03.4 molecular cloud is 918$^{+15}_{-{18}}$. 

\section{Result and discussion}   
\subsection{Distances of the molecular clouds}
\begin{figure*}
    \centering
	\includegraphics[width=1\linewidth]{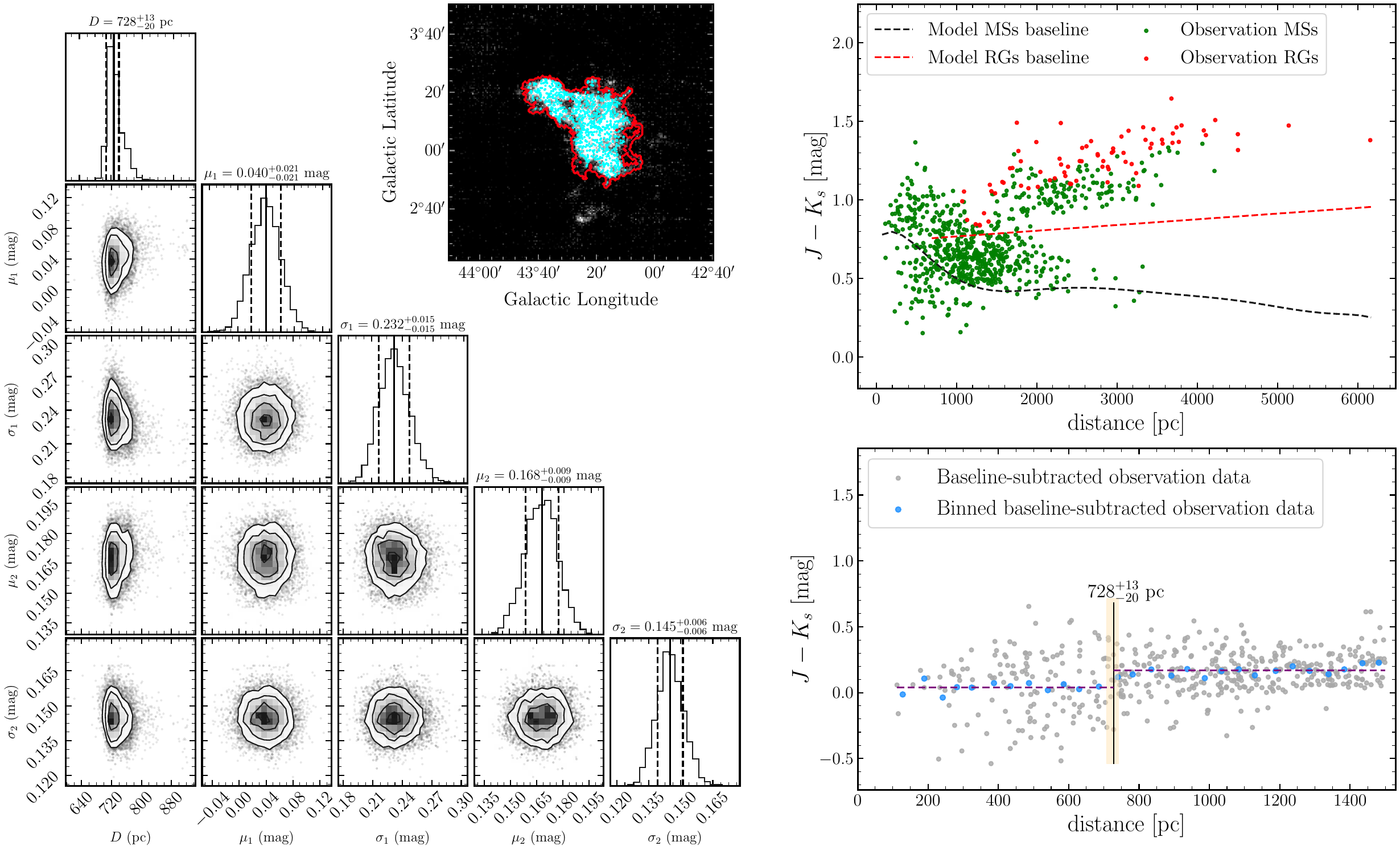}
	\caption{Distance of G043.3+03.1. The red contours refer to the molecular cloud footprint, and the cyan points represent the on-cloud stars. The left and bottom right panels are corner maps and estimate the distance of the MCMC sampling (see Fig.~\ref{distance-mcmc} for details). The red and green points in the top right panel present RGs and MSs that were distinguished in the observation data, respectively. The dashed black and red lines show the model-fitted RGs and MSs baseline, respectively.}
	\label{117135-distance}
\end{figure*}

With the above new method, we derived the distances of 27 molecular clouds. The results are summarized in Table \ref{Table1}. The systematic error is not included. From left to right, we display the molecular cloud name (1), the ratio of the distance errors to the distance (2), the minimum CO emission of the on-cloud stars (3), the maximum distances to the on-cloud stars (4), the number of the on-cloud stars (5), the estimated distances based on the TRILEGAL model (6), the estimated distances based on the GALAXIA model (7), and the distances of \citet{Yan+2021} (8). The nearest distance to these molecular clouds is 315 pc, and the farthest distance is 2730 pc. The distance uncertainties might be larger for distant molecular clouds, such as G032.9-01.8 and G042.9+04.4. The results for G045.1-03.4 and G043.3+03.1 are displayed in Fig.~\ref{distance-mcmc} and \ref{117135-distance}, and the other 25 molecular clouds are illustrated in the online supplementary material \footnote{\url{https://zenodo.org/records/10696502}}.

\begin{table*}
    \centering
    \renewcommand\arraystretch{1.55}  	
	\centering
	\setlength{\tabcolsep}{2.8mm}{   
    \caption{Distances of the molecular clouds.}
    \label{Table1}
    \begin{tabular}{cccccccc}
        \hline  
        Name & $ V_{\rm LSR}$ & CO$_{\rm cut}$ $^{\textcolor{blue}{1}}$ & $D_{\rm cut}$ & N$_{\rm stars}$ $^{\textcolor{blue}{2}}$ & $D_{\rm TRI}$ $^{\textcolor{blue}{3}}$ & $D_{\rm GAL}$ $^{\textcolor{blue}{4}}$ & $D_{\rm Lit}$ $^{\textcolor{blue}{5}}$     \\
         & (km s$^{-1}$) & (K km s$^{-1}$) & (pc) &  & (pc) & (pc) & (pc) \\
        (1) & (2) & (3) & (4) & (5) & (6) & (7) & (8) \\
        \hline
        G$026.9-03.5$	&	16.3 	&	2.5 	&	1000	&	263	&	564	$^{+20}_{-{26}}$	&	559	$^{+26}_{-{21}}$	&	445	$^{+17}_{-{17}}$	\\
        G$027.8-02.1$	&	18.0 	&	4.2 	&	1000	&	1876	&	553	$^{+7}_{-{7}}$	&	473	$^{+3}_{-{4}}$	&	484	$^{+8}_{-{8}}$	\\
        G$028.6-02.9$	&	18.5 	&	0.9 	&	1500	&	77	&	592	$^{+32}_{-{57}}$	&	594	$^{+28}_{-{50}}$	&	826	$^{+94}_{-{94}}$	\\
        G$032.9-01.8$	&	22.4 	&	0.8 	&	4000	&	113	&	2730	$^{+130}_{-{118}}$	&	2698	$^{+93}_{-{88}}$	&	2392	$^{+192}_{-{192}}$	\\
        G$034.2-01.7$	&	6.6 	&	2.0 	&	1500	&	244	&	682	$^{+21}_{-{15}}$	&	579	$^{+27}_{-{27}}$	&	532	$^{+61}_{-{61}}$	\\
        G$037.0+04.6$	&	22.8 	&	1.2 	&	2000	&	164	&	1181	$^{+50}_{-{38}}$	&	1220	$^{+65}_{-{71}}$	&	1016	$^{+85}_{-{85}}$	\\
        G$038.1-01.2$	&	7.7 	&	2.0 	&	2000	&	301	&	842	$^{+23}_{-{19}}$	&			&	698	$^{+64}_{-{64}}$	\\
        G$038.4-03.3$	&	7.5 	&	1.5 	&	2000	&	302	&	877	$^{+48}_{-{60}}$	&			&	872	$^{+66}_{-{66}}$	\\
        G$039.4-02.7$	&	15.0 	&	2.5 	&	1500	&	302	&	684	$^{+53}_{-{49}}$	&	688	$^{+56}_{-{53}}$	&	493	$^{+20}_{-{20}}$	\\
        G$041.5+02.3$	&	17.6 	&	4.9 	&	1500	&	1707	&	919	$^{+11}_{-{10}}$	&	928	$^{+12}_{-{13}}$	&	909	$^{+13}_{-{13}}$	\\
        G$041.5-01.8$	&	16.8 	&	1.2 	&	1500	&	304	&	806	$^{+24}_{-{44}}$	&	711	$^{+59}_{-{58}}$	&	708	$^{+65}_{-{65}}$	\\
        G$041.9+02.2$	&	5.4 	&	1.5 	&	800	&	1567	&	315	$^{+8}_{-{7}}$	&	240	$^{+7}_{-{6}}$	&	314	$^{+15}_{-{15}}$	\\
        G$042.1-02.1$	&	23.4 	&	1.2 	&	2000	&	335	&	866	$^{+45}_{-{59}}$	&			&	1066	$^{+55}_{-{55}}$	\\
        G$042.9+04.4$	&	22.7 	&	3.0 	&	3500	&	49	&	1652	$^{+80}_{-{110}}$	&	1625	$^{+100}_{-{140}}$	&	1441	$^{+68}_{-{68}}$	\\
        G$043.3+03.1$	&	10.2 	&	2.5 	&	1500	&	485	&	728	$^{+13}_{-{20}}$	&	727	$^{+16}_{-{18}}$	&	723	$^{+16}_{-{16}}$	\\
        G$043.4-04.2$	&	26.1 	&	1.2 	&	2000	&	102	&	910	$^{+46}_{-{49}}$	&	918	$^{+38}_{-{57}}$	&	1021	$^{+52}_{-{52}}$	\\
        G$044.5+02.7$	&	14.6 	&	2.0 	&	1500	&	350	&	662	$^{+32}_{-{31}}$	&	649	$^{+68}_{-{19}}$	&	621	$^{+16}_{-{16}}$	\\
        G$044.7-04.2$	&	22.8 	&	0.9 	&	2000	&	104	&	932	$^{+60}_{-{123}}$	&			&	1154	$^{+166}_{-{166}}$	\\
        G$044.7+04.0$	&	20.0 	&	3.0 	&	1500	&	1643	&	861	$^{+6}_{-{7}}$	&	860	$^{+7}_{-{7}}$	&	829	$^{+13}_{-{13}}$	\\
        G$045.1-03.4$	&	20.7 	&	1.2 	&	2000	&	196	&	918	$^{+15}_{-{18}}$	&	923	$^{+20}_{-{23}}$	&	923	$^{+72}_{-{72}}$	\\
        G$045.4-04.2$	&	18.8 	&	2.5 	&	1500	&	851	&	894	$^{+13}_{-{13}}$	&	899	$^{+14}_{-{13}}$	&	931	$^{+25}_{-{25}}$	\\
        G$045.5-03.3$	&	20.2 	&	2.0 	&	1500	&	133	&	937	$^{+12}_{-{17}}$	&	940	$^{+14}_{-{18}}$	&	873	$^{+41}_{-{41}}$	\\
        G$046.9-04.8$	&	15.7 	&	0.9 	&	1500	&	99	&	847	$^{+53}_{-{54}}$	&	822	$^{+76}_{-{44}}$	&	841	$^{+45}_{-{45}}$	\\
        G$046.9+01.7$	&	6.8 	&	2.8 	&	1000	&	298	&	426	$^{+18}_{-{26}}$	&			&	491	$^{+103}_{-{103}}$	\\
        G$047.3-03.1$	&	18.7 	&	2.0 	&	2000	&	177	&	742	$^{+23}_{-{39}}$	&	751	$^{+24}_{-{46}}$	&	912	$^{+99}_{-{99}}$	\\
        G$047.3-03.0$	&	24.4 	&	1.5 	&	2000	&	164	&	1084	$^{+45}_{-{49}}$	&			&	1127	$^{+88}_{-{88}}$	\\
        G$049.4-04.6$	&	18.7 	&	0.9 	&	2000	&	81	&	851	$^{+33}_{-{42}}$	&	855	$^{+34}_{-{46}}$	&	957	$^{+96}_{-{96}}$	\\
        \hline
        \multicolumn{3}{l}{$^1$ The lower threshold of the CO emission for on-cloud stars.}\\
        \multicolumn{3}{l}{$^2$ The number of the on-cloud stars used to calculate the distance.}\\
        \multicolumn{3}{l}{$^3$ The estimated distances based on the TRILEGAL model.}\\
        \multicolumn{3}{l}{$^4$ The estimated distances based on the GALAXIA model.}\\
        \multicolumn{3}{l}{$^5$ The distances of \citet{Yan+2021}.}\\
    \end{tabular}}
\end{table*}

\subsection{Effect of the parameter choice}
\begin{figure*}
    \centering
	\includegraphics[width=0.96\linewidth]{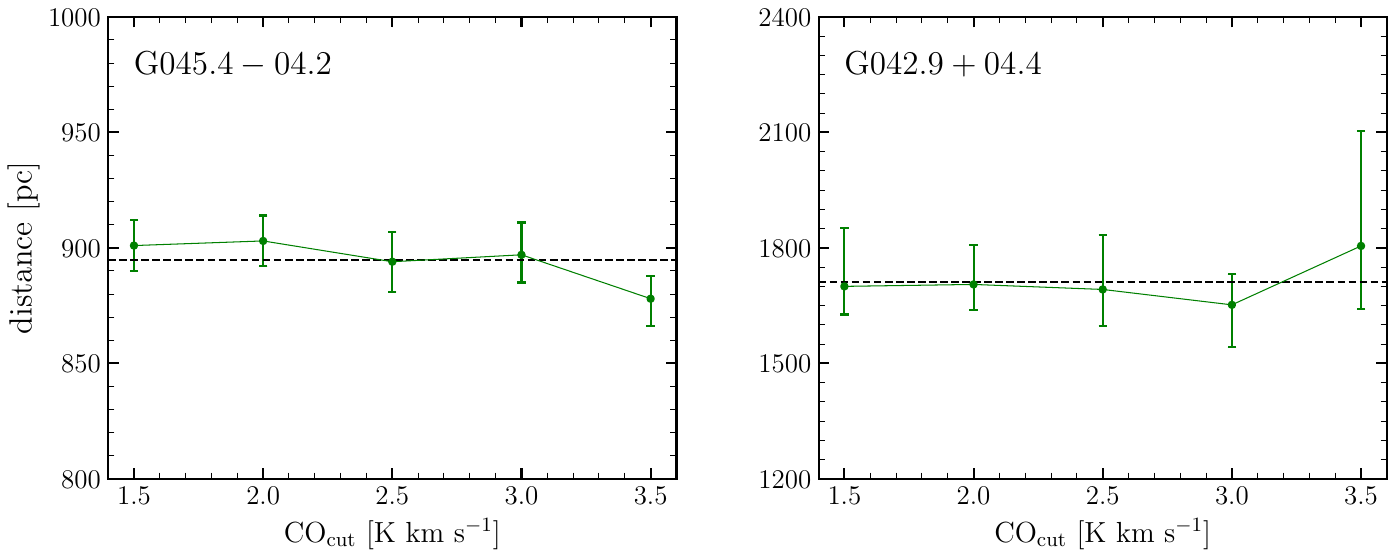}
	\caption{Molecular cloud distances and uncertainties after changing the CO$_{\rm cut}$ parameter. The dashed black line is the average of the multiple distances. The left and right panels show the effect of CO$_{\rm cut}$ in G045.4-04.2 and G042.9+04.4, respectively. The range of CO$_{\rm cut}$ is 3$\sigma_{\rm int}$ to 7$\sigma_{\rm int}$, where both clouds $\sigma_{\rm int}$ are 0.5.}
	\label{parameter-effect}
\end{figure*}

We explored the effect of the method dependence parameters. In the selection of the on-cloud stars, we involved the CO$_{\rm cut}$ parameter above which on-cloud stars are selected. The choice of CO$_{\rm cut}$ may affect the determination of the distance. We examined the CO$_{\rm cut}$ parameter. To explore the influence of this parameter, we only changed the CO$_{\rm cut}$ and left other parameters unchanged, and we compared the molecular cloud distance variations. As shown in Fig.~\ref{parameter-effect}, we display the molecular cloud distances and uncertainties after changing CO$_{\rm cut}$, and the range of CO$_{\rm cut}$ is taken as 3$\sigma_{\rm int}$ $\sim$ 7$\sigma_{\rm int}$. The distance uncertainty of G042.9+04.4 is larger than that of G045.4-04.2 because the fewer background stars of the far-distant molecular cloud result in a larger distance uncertainty. There are slight fluctuations in the molecular cloud distances and uncertainties, but they are within reasonable limits. In general, CO$_{\rm cut}$ has only a weak effect on the distance estimate. Thus, the new method does not depend strongly on the boundary of the cloud. 

\subsection{Uncertainties in the cloud distance}
After solving the Bayesian model by MCMC sampling, we estimated reliable statistical uncertainties from the posterior distribution of the model parameters. However, we considered additional uncertainties in the method. There are mainly the following points.

(1) Uncertainties in the distance and color of the observation data. The systematic uncertainty of the  $J-K{_s}$ color is much smaller than that of the distance, and therefore, we mainly considered the distance error. The adopted stellar distances were estimated by \citet{BailerJ+2021}, who transferred the \textit{Gaia} EDR3 parallaxes to distances using a Bayesian approach. \citet{Lindegren+2018,Lindegren+2021} reported that the systematic errors of the \textit{Gaia} DR2 parallaxes are approximately 0.1 mas and that those of the \textit{Gaia} EDR3 parallaxes are even smaller. Therefore, a molecular cloud is at a distance of 1000 pc (1 mas), so that the systematic uncertainty of the distance caused by the system parallax error is about 10 $\%$ at most. 

(2) Uncertainties in the Milky Way model. To calibrate the color distribution, we used stellar populations generated with the TRILEGAL model to fit the baseline. We caused the model data to match the observation data, as is possible in the off-cloud region, by using the same photometric system and adjusting the model parameters. Since the fitted baseline from the model data may deviate, it could potentially affect the distance determination. In addition, RGs and reddened MSs are distinguished based on the model probability, which may also cause a distance uncertainty. As a comparison, we used another Milky Way model, the GALAXIA model \citep{Sharma+2011}. With similar parameters as the TRILEGAL model, the GALAXIA synthetic population matches the observation data well. However, the red dwarf branch of the GALAXIA synthetic populations is less pronounced in some clouds, which causes the baseline fit by the GALAXIA synthetic population to be incapable of calibrating the observation data at a distance of $<$ 1 kpc. Therefore, we measured the distances of 21 clouds using the GALAXIA model. Figure~\ref{distance-comparison} depicts the good consistency between the distances of the 21 clouds estimated from the TRILEGAL and GALAXIA models in the left panel. The choice of a different galaxy model affects the distance estimates of clouds, depending on whether the chosen galaxy model can match the observed stellar populations in the color-distance diagram. The TRILEGAL and GALAXIA models can both simulate synthesized stellar populations that are consistent with the observations. The distances estimated from the two models match each other very well, suggesting that the method proposed in this work does not depend too strongly on the model. 

(3) Uncertainties in the Bayesian model. Our Bayesian model is simple. It assumes that the molecular cloud is at a single distance and that the foreground and background stellar distributions obey a Gaussian distribution. A violation of these assumptions would lead to distance errors.

\subsection{Comparison with previous results}
\begin{figure*}
    \centering
    \includegraphics[width=0.49\linewidth]{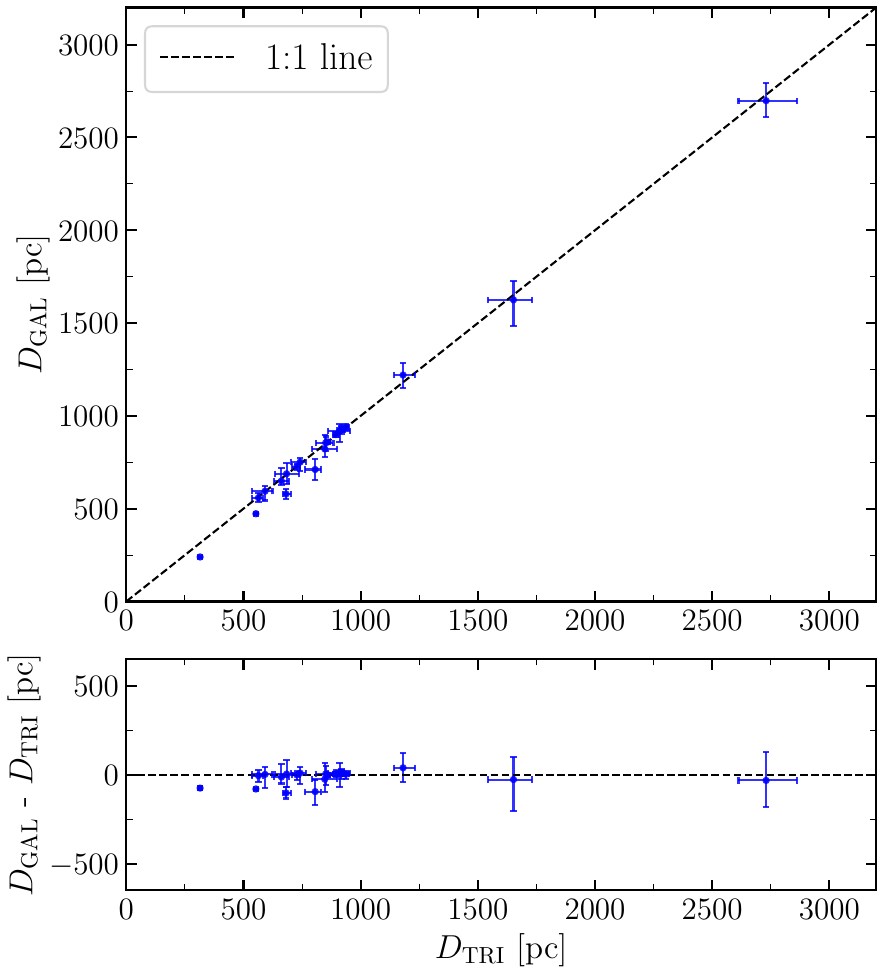}
    \includegraphics[width=0.49\linewidth]{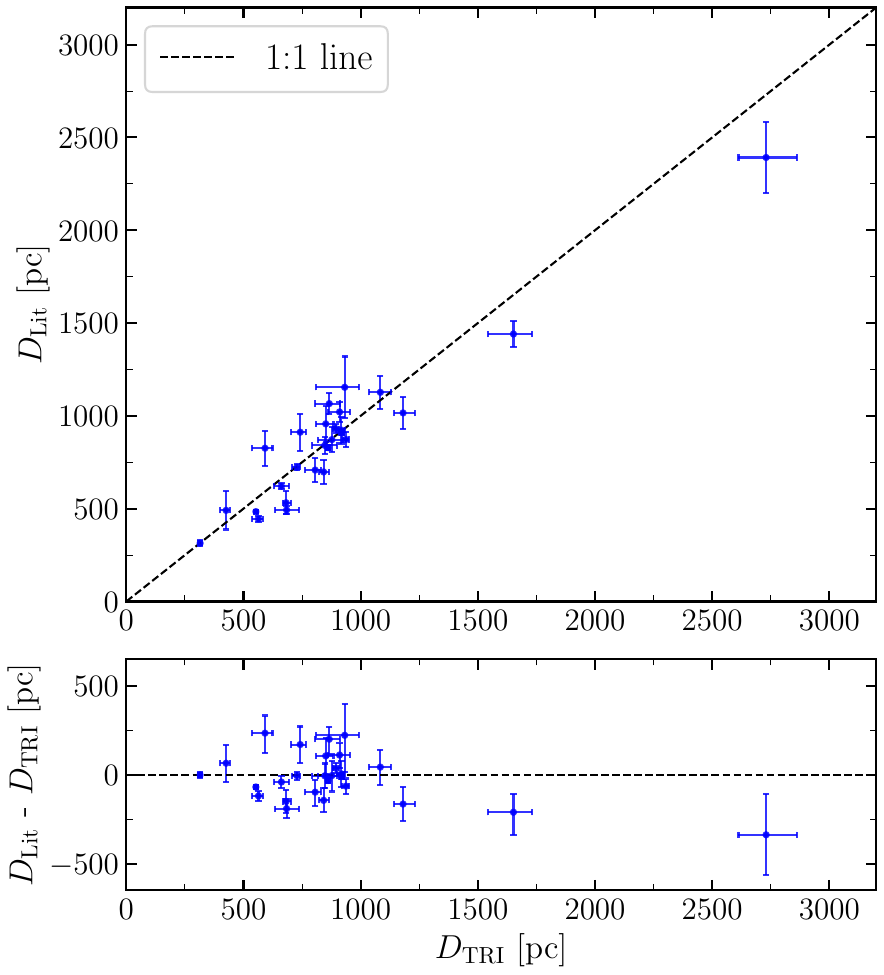}
     \caption{Left panels: Comparison between the distances of 21 clouds estimated from the TRILEGAL ($D_{\rm TRI}$) and GALAXIA ($D_{\rm GAL}$) models. Right panels: Comparison between the distances of the 27 clouds estimated with the TRILEGAL model ($D_{\rm TRI}$) and those ($D_{\rm Lit}$) of \protect\citet{Yan+2021}. }
	\label{distance-comparison}
\end{figure*}

We compared the distances we derived from the current method with the results of \citet{Yan+2021}. The relation between our distances and the extinction distances by \citet{Yan+2021} is shown in the right panels of Fig.~\ref{distance-comparison}. The two methods agree well, especially the estimated distance within 1.5 kpc. Across the range of distances explored, the typical scatter between our distance compared to the extinction distance is about 10\%. The distance estimates from TRILEGAL for the two clouds at distances $>1.5\,\mathrm{kpc}$ seem to be systematically larger than the values of \citet{Yan+2021}. Nevertheless, the TRILEGAL and GALAXIA models return very close estimates of the distances of the two clouds. On the other hand, only two clouds have distances farther than $\sim 1.5\,\mathrm{kpc}$. It is hard to say whether our method derives a systematic distance bias for these distant clouds in comparison to the values of \citet{Yan+2021}. We only adopted on-cloud stars, which calibrate the color distribution using the TRILEGAL stellar population. This greatly reduces the requirement for an accurate determination of the molecular cloud boundaries.

As indicated in Fig.~\ref{117135-distance}, we display the distance of G043.3+03.1. Our estimated distances of G043.3+03.1 is 728$^{+13}_{-{20}}$, and we compared the derived distance with the result from the literature. The literature estimate of the distance for G043.3+03.1 determined by BEEP is 731$^{+40}_{-{38}}$ \citep[c.f. fig.7(b) in][]{Yan+2020}, which is consistent with our current result. The dispersion of the differences between our distance and that from the literature is minor.

\subsection{Limitations of our method}
In the method, we calibrated the color of the on-cloud stars and considered the influence of red giant stars on the color jump, but there are still some limitations in calculating the molecular cloud distance, as follows: First, it is difficult to measure molecular clouds at too great distances ($>$3kpc) because the data are limited and have large errors. Second, the number of on-cloud stars is not enough, which may mean that the distance cannot be measured. Third, the reddened stars caused by molecular clouds are not sufficient to make an obvious color jump. In this case, the distance measured may not be accurate. Fourth, when two or more molecular clouds overlap along the LOS, multiple jumps become possible. Our method currently does not solve the multijump problem. This problem may be solved in the future.

\section{Summary}  
We presented a new method for estimating the distances to molecular clouds. As the trial work to evaluate the method, we estimated the distances for 27 molecular clouds using $J-K{_s}$ colors and the distances provided by 2MASS and \textit{Gaia} EDR3. The measured clouds were selected from \citet{Yan+2021}. For each molecular cloud, we selected on-cloud stars based on the cloud boundary, defined by a certain threshold from the CO map of the MWISP survey. We adopted the stellar populations generated with the TRILEGAL galaxy model as the baseline instead of off-cloud stars. Considering that RGs may affect the distance determination, we distinguished between RGs and MSs for the model and observation data, respectively. The colors of the observation data were calibrated using the corresponding model-fit baseline. Based on the baseline-subtracted data, we successfully used Bayesian analysis and MCMC sampling to estimate the distances of the selected clouds, ranging from $\sim$315 to $\sim$2730 pc. By replacing TRILEGAL with the GALAXIA galaxy model, we measured the distances of 21 clouds using GALAXIA, which are consistent with the distances estimated by TRILEGAL. The typical statistical uncertainties of the distances are $\sim$ 5$\%$, and the systematic uncertainties from the method are $\sim$ 10$\%$. In the future, we will measure more distances of molecular clouds from the MWISP survey, such as complex regions with indistinguishable molecular cloud boundaries.

\begin{acknowledgements}
This work was supported by the National Natural Science Foundation of China (grants Nos.U2031202,12373030). ZC acknowledges the Natural Science Foundation of Jiangsu Province (grants No. BK20231509). This work make use of the data from the Milky Way Imaging Scroll Painting (MWISP) project, which is a multi-line survey in $^{12}$CO/$^{13}$CO/C$^{18}$O along the northern galactic plane with PMO-13.7m telescope. We are grateful to all the members of the MWISP working group, particularly the staff members at the PMO 13.7 m telescope, for their long-term support. MWISP is sponsored by the National Key R\&D Program of China with grants 2023YFA1608000 and 2017YFA0402701, and the CAS Key Research Program of Frontier Sciences with grant QYZDJ-SSW-SLH047. This work has made use of data from the European Space Agency (ESA) mission Gaia (\url{https://www.cosmos.esa.int/gaia}), processed by the Gaia Data Processing and Analysis Consortium (DPAC, \url{https://www.cosmos.esa.int/web/gaia/dpac/consortium}). Funding for the DPAC has been provided by national institutions, in particular the institutions participating in the Gaia Multilateral Agreement. This publication makes use of data products from the Two Micron All Sky Survey, which is a joint project of the University of Massachusetts and the Infrared Processing and Analysis Center/California Institute of Technology, funded by the National Aeronautics and Space Administration and the National Science Foundation.
\end{acknowledgements}

%
\bibliographystyle{aa} 
\bibliography{aa} 
%

\end{CJK*}
\end{document}